\documentclass[12pt]{article}
\usepackage{scicite}

\newenvironment{sciabstract}{%
\begin{quote} \bf}
{\end{quote}}

\usepackage{times}
\usepackage{url}
\usepackage{tikz}
\usetikzlibrary{positioning}
\usetikzlibrary{calc}
\usepackage{pgfplots, pgfplotstable}
\pgfplotsset{compat=1.14}

\newcommand{\etal}{\textit{et al}.}

\newcommand{\mps}[1]{}
\newcommand{\ks}[1]{}
\newcommand{\cwh}[1]{}

\title{The Causal Link between News Framing and Legislation}

\author
{Karthik Sheshadri, Chung-Wei Hang, Munindar P.~Singh$^{1\ast}$\\
\\
\normalsize{$^{1}$Department of Computer Science, North Carolina State University}\\
\\
\normalsize{$^\ast$E-mail: kshesha@ncsu.edu}
}

\date{}

\begin{document} 


\baselineskip24pt


\maketitle


\begin{sciabstract}
  We demonstrate that framing, a subjective aspect of news, is a causal precursor to both significant public perception changes, and to federal legislation. We posit, counter-intuitively, that topic news volume and mean article similarity increase and decrease together. We show that specific features of news, such as publishing volume , are predictive of both sustained public attention, measured by annual Google trend data, and federal legislation. We observe that public attention changes are driven primarily by periods of high news volume and mean similarity, which we call \emph{prenatal periods}.  Finally, we demonstrate that framing during prenatal periods may be characterized by high-utility news \emph{keywords}. 
\end{sciabstract}



The effect of news on public behavior  has been the subject of considerable scientific interest. Prior work has established that news framing influences public perception \cite{gunther1998persuasive,mutz1997reading}, affects technology development \cite{hoadley2010privacy,quora}, and contributes to setting agendas \cite{agenda}. Most recently, publishing from small news outlets has been shown to increase short-term public involvement in specific domains \cite{king}.


Our work enhances existing understanding by explicitly modeling the Granger  causal \cite{granger} link between specific news characteristics, public opinion, and federal legislation. Firstly, we demonstrate a direct predictive relationship between news characteristics and federal legislation, independently of public opinion changes. Secondly, we show that the public reacts predictably to high-volume, high-similarity news periods, but does not otherwise react. We note that King {\etal}'s \cite{king} approach artificially created such news conditions for short time periods and reader subsets.

Thirdly, news reporting in general introduces subjective biases, collectively referred to as framing \cite{entman}. Whereas news publishing is ordinarily event driven, we demonstrate that high volume and similarity news periods can occur spontaneously, without event-based drivers, as a Granger causal effect of news framing. Consistent with our result on prenatal periods, such periods brought about by framing are equally influential in predicting public approval (the fraction of the public that approves or disapproves of a particular position) and legislation. This finding demonstrates that news framing is as influential as the actual events and facts that the news reports on. 


We also demonstrate that news publishing volume within a specific domain is a reliable long-term predictor of public attention (the number of people who demonstrated interest in a domain by conducting an Internet search), measured annually using Google Trend data.


The causal flow we discovered is depicted in Fig~\ref{fig:influence}. We confirmed each link using a directional Granger causality test, which evaluates the influence of a causal time series on a dependent one. Our choice of Granger causality over a structural model was deliberate, since we wished to infer rather than assume structure and direction. The Granger

Our approach is similar in spirit to King {\etal} \cite{king}, but yields several novel results. We show that news is a Granger causal precursor  to legislation. Our analysis applies to a larger population than the outlets used in \cite{king}, since the NYT and Guardian enjoy wide circulation and readership. We measure public perception annually rather than over a period of weeks, as King {\etal} do. We distinguish between fact-based reporting and framing, and demonstrate that framing in itself is a causal predictor of public attention and legislation. 

We used a set of online news Application Programming Interfaces (APIs), including the New York Times \cite{nytapi} and the Guardian \cite{guardianapi} to create our news datasets. We began with an initial keyword search for articles in each domain of interest, and then applied multiple rounds of supervised classification to weed out irrelevant articles, resulting in a dataset with a median precision of 0.93 and a median per domain inter-rater agreement index (Kappa) \cite{vg05} of 0.87. Our classification relied on ground truth data obtained from multiple raters, employing a scheme whereby an article was considered as belonging to a domain if and only if the article could not have been published with the domain component removed. While we did not directly measure recall, since news publications have a strong incentive to broadly cover events, and because the NYT and the Guardian have the largest and fifth largest circulations in America and the world respectively, we expect that our datasets have high recall. 


Our observations stem from a remarkable pattern that holds across nearly a hundred thousand articles from about fifty domains. In each domain, the number of news articles published in a certain year varies directly with mean corpus similarity, with significant Granger causal F-measures in both directions. This finding is surprising, since one would expect a larger volume of articles to discuss a larger variety of subjects. Instead, we found that domain news publishing volume is mainly event driven, and events increase not only the mean similarity of the corpus but also its \emph{volume}, for example, the number of Surveillance articles increased by 220\% in 2013, with 93\% of the total being primarily about Snowden. It is well known that news is event driven, however, the discovery of a causal relationship between article volume and mean corpus similarity is a novel finding of our work. In order to estimate mean similarity, we generate all $n\choose2$ pairs of articles from a corpus of size $n$, and use their average cosine distance \cite{cd} in Doc2Vec \cite{doc} space as our similarity metric.

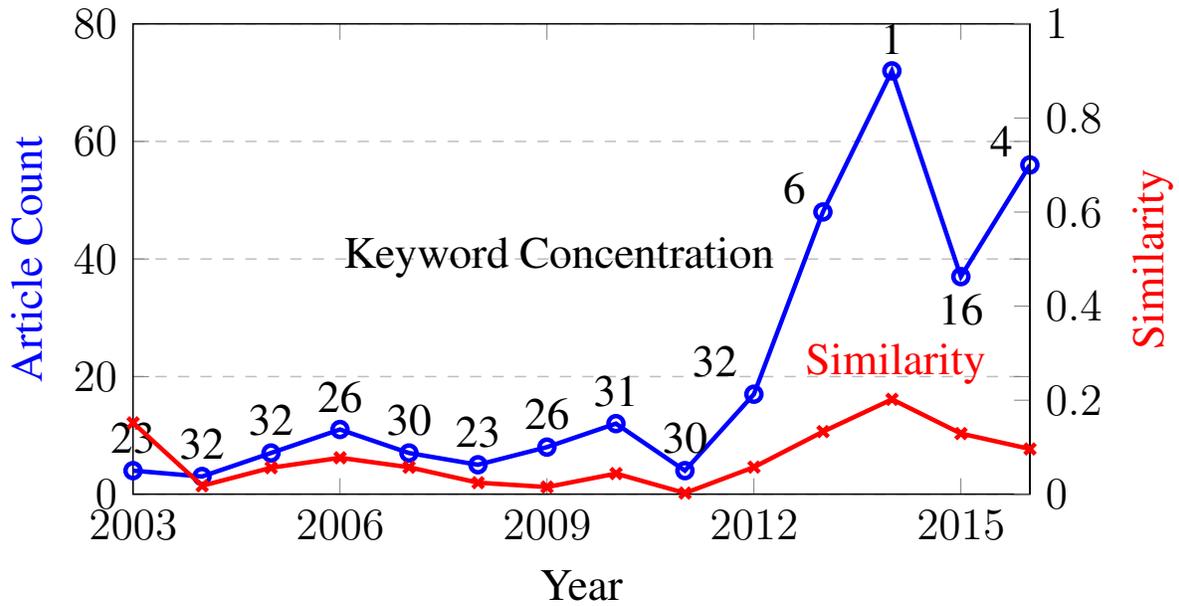
\begin{figure}[!htb]
    \centering
    \resizebox{0.99\columnwidth}{!}{%
\begin{tikzpicture}
    \begin{axis}[
        title={},
        height=6cm,
        width=10cm,
        xlabel={Year},
        ylabel={\textcolor{blue}{Article Count}},
        xmin=2003, xmax=2016,
        ymin=0, ymax=80,
        xtick={2003,2006,2009,2012,2015},
        ytick={0,20,40,60,80,100},
        legend pos=north west,
        legend style={font=\tiny},
        mark size=2.0pt,
        ymajorgrids=true,
        grid style=dashed,
        xticklabel style={/pgf/number format/.cd, set thousands separator={}},
    ]

    \addplot[
        blue,
        very thick,
        mark=o,
        mark options={fill=white},
        visualization depends on=\thisrow{alignment} \as \alignment,
        nodes near coords, 
        point meta=explicit symbolic, 
        every node near coord/.style={anchor=\alignment} 
    ] table [
        meta index=2, 
        row sep=crcr
    ] {
x       y   label   alignment\\
2003    4   \textcolor{black}{23}      -90\\
2004    3   \textcolor{black}{32}      -90\\
2005    7   \textcolor{black}{32}      -90\\
2006   11   \textcolor{black}{26}      -90\\
2007    7   \textcolor{black}{30}      -90\\
2008    5   \textcolor{black}{23}      -90\\
2009    8   \textcolor{black}{26}      -90\\
2010    12  \textcolor{black}{31}      -90\\
2011    4   \textcolor{black}{30}      -90\\
2012    17  \textcolor{black}{32}      -40\\
2013    48  \textcolor{black}{6}       -40\\
2014    72  \textcolor{black}{1}       -90\\
2015    37  \textcolor{black}{16}      90\\
2016    56  \textcolor{black}{4}       -40\\
} node[above left,pos=0.4] {\textcolor{black}{Keyword Concentration}};



    \end{axis}

   \begin{axis}[
        height=6cm,
        width=10cm,
        axis y line*=right,
        axis x line=none,
        xmin=2003, xmax=2016,
        ymin=0, ymax=1,
        ytick={0,0.20,0.40,0.60,0.80,1.00},
        ylabel={\textcolor{red}{Similarity}},
        ylabel near ticks,
        yticklabel pos=right
    ]

  \addplot [very thick,red,mark=x]
        coordinates {(2003,0.151819195) (2004,0.01804966519) (2005,0.0564282172339713) (2006,0.0775952248754797) (2007,0.0574131362074208) (2008,0.0246369129971187) (2009,0.0153641043316717) (2010,0.0440065159756945) (2011,0.00257047230208002) (2012,0.0577160220712493) (2013,0.1332121718) (2014,0.2018622475) (2015,0.1290804976) (2016,0.0963360531614937)}
        node[above,pos=0.85] {Similarity};

    \end{axis}

\end{tikzpicture}%
}
    \caption{News volume, correlation, and sentiment as predictors of legislation in Surveillance.}
    \label{fig-surveillance}
\end{figure}

Thence, we posit the idea of a prenatal period of domain news, which is characterized by simultaneously high article volume and mean similarity. We studied public reaction to news and found that causal changes in public attention and approval occurred \emph{only} after such periods, establishing a prenatal period as the necessary and sufficient condition for such changes. Fig.~\ref{fig-surveillance} illustrates a prenatal period of the Surveillance domain. 

\begin{figure}[!htb]
    \centering
    \resizebox{0.99\columnwidth}{!}{%
\begin{tikzpicture}
        \begin{axis}[
        title={},
        height=6cm,
        width=10cm,
        xlabel={Year},
        ylabel={Article and Law Count},
        xmin=1991, xmax=2016,
        ymin=0, ymax=200,
        xtick={1990,1995,2000,2005,2010,2015},
        ytick={0,30,60,90,120,150,180},
        legend pos=north west,
        legend style={font=\tiny},
        mark size=2.0pt,
        ymajorgrids=true,
        grid style=dashed,
        xticklabel style={/pgf/number format/.cd, set thousands separator={}},
        ]

        \addplot [very thick,blue,mark=o]
        coordinates {(1991,30) (1992,24) (1993,39) (1994,30) (1995,23) (1996,24) (1997,42) (1998,54) (1999,67) (2000,142) (2001,165) (2002,121) (2003,162) (2004,166) (2005,134) (2006,177) (2007,85) (2008,67) (2009,34) (2010,41) (2011,36) (2012,136) (2013,103) (2014,87) (2015,66) (2016,56)}
        node[left=.6cm, pos=0.3] {Articles};
        
       \addplot [very thick,brown,mark=square]
        coordinates {(1991,0) (1992,0) (1993,0) (1994,0) (1995,0) (1996,0) (1997,0) (1998,0) (1999,0) (2000,100) (2001,100) (2002,100) (2003,100) (2004,100) (2005,100) (2006,100) (2007,0) (2008,0) (2009,0) (2010,0) (2011,0) (2012,0) (2013,100) (2014,100) (2015,0) (2016,0)}
        node[below=.6cm, pos=0.264] {COPPA};

    \end{axis}

\begin{axis}[
    height=6cm,
        width=10cm,
  axis y line*=right,
  axis x line=none,
xmin=1991, xmax=2016,
  ymin=0, ymax=1,
ytick={0,0.20,0.40,0.60,0.80,1.00},
  ylabel={GT Index}     
]

           \addplot [very thick,black,mark=square]
        coordinates {(2005,.853) (2006,.692) (2007,.609) (2008,.526) (2009,.534) (2010,.513) (2011,.547) (2012,.611) (2013,.592) (2014,.659) (2015,.586) (2016,.567)}
        node[above=.5cm,pos=0.5] {GT Index};

\addplot [very thick,red,mark=x]
        coordinates {(1991,0.494891730377283/2) (1992,0.306500824918334/2) (1993,0.450524639496613/2) (1994,0.244670987124601/2) (1995,0.500434273364260/2) (1996,0.167088600014157/2) (1997,0.234390727887453/2) (1998,0.313505617206397/2) (1999,0.636443234807432/2) (2000,0.950829151665395/2) (2001,0.836356036667068/2) (2002,0.524935759360227/2) (2003,0.9234556/2) (2004,0.9502805248098893/2) (2005,0.672068049974419/2) (2006,0.778970431475917/2) (2007,0.593680956199340/2) (2008,0.538252502878553/2) (2009,0.243932145909111/2) (2010,0.403037597968340/2) (2011,0.365651526803978/2) (2012,0.798736564775931/2) (2013,0.775975894472513/2) (2014,0.730283954635817/2) (2015,0.340809422941126/2) (2016,0.466103653044484/2)}
        node[above=0.1cm, pos=0.15] {Similarity};

    \end{axis}

\end{tikzpicture}%
}
    \caption{News and legislation}
    \label{fig-coppa}
\end{figure}
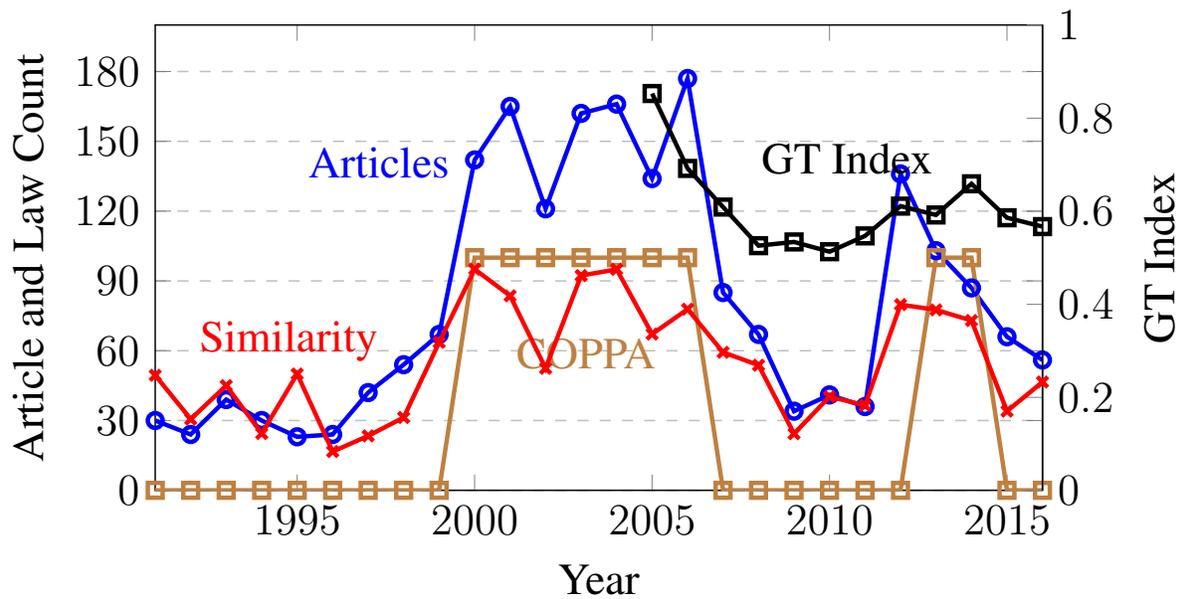

To establish the causal effect of news on legislation, we considered all federal legislation enacted beginning from the $101^{st}$ United States Congress until the (present) $116^{th}$ Congress. Our choice was motivated by the fact that our APIs provide access to data beginning in 1991. Approximately 42\% of legislation was foreshadowed by a prenatal domain news period. Note that while we do not claim prenatal news periods to be a necessary condition for legislation, we found that the probability of legislation succeeding a prenatal period was causally significant. We illustrate our approach and results in Fig~\ref{fig-coppa}, using a compelling example from the domain of Child Privacy. The primary laws governing children's privacy protection in the United States are COPPA \cite{coppa} and FERPA \cite{ferpa}. COPPA was originally introduced in April
1998, and went through a series of amendments from 1999 through 2005, and again from 2012--2013. FERPA was enacted in 1974. Due to the unavailability of children's privacy news articles before 1974 (a keyword search in the NYT developers API returns zero articles), we restrict our analysis to COPPA. The causal variables of interest in Fig~\ref{fig-coppa} are annual news volume (blue), and mean pairwise article similarity (red).  Observe that the number of news articles published on the topic doubled between 1995--1999, coupled with a simultaneous increase in mean article similarity. Following this prenatal period, COPPA legislation was promulgated through the period ending in 2005. Another prenatal period occurs before the revival of COPPA amendments in 2012.

We tested the Granger causal flow depicted in Fig.~\ref{fig:influence} over the set of domains obtained earlier, yielding a median F-measure and critical value \mps{never defined} of 5.63 and 4.45, respectively, at the 0.05 level. 


Google Trends (GT) \cite{gtapi} estimate public interest in a topic of interest by measuring related searches worldwide over chosen time periods. Since 79\% of US and UK residents use the Internet and 74\% use Google as their primary search, GT is a representative measure of public attention. We collected news data from over 20 domains, and modeled Granger causality tests between article volume and GT volume, yielding a median F-measure and critical value of 5.72 and 4.39, respectively, at the 0.05 level. \mps{Somewhere we should define exactly what the Granger test is applied on. We should add an appendix with formal definitions. } \ks{methods are to be described in the supplementary material.}\mps{Describe within the paper what the Granger test is on: a time series of \ldots (in English), for at least one of them}

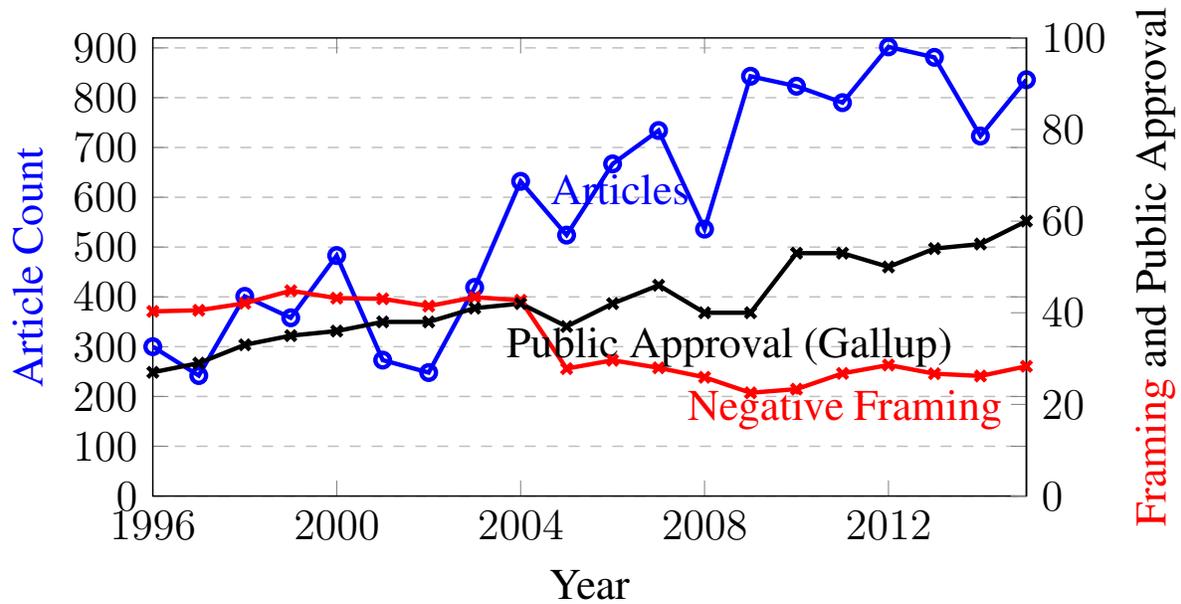
\begin{figure}[!htb]
    \centering
    \resizebox{0.99\columnwidth}{!}{%
\begin{tikzpicture}
    \begin{axis}[
        title={},
        height=6cm,
        width=10cm,
        xlabel={Year},
        ylabel={\textcolor{blue}{Article Count}},
        xmin=1996, xmax=2015,
        ymin=0, ymax=920,
        xtick={1996,2000,2004,2008,2012,2016},
        ytick={0,100,200,300,400,500,600,700,800,900},
        legend pos=north west,
        legend style={font=\tiny},
        mark size=2.0pt,
        ymajorgrids=true,
        grid style=dashed,
        xticklabel style={/pgf/number format/.cd, set thousands separator={}},
    ]

    \addplot [very thick,blue,mark=o]
        coordinates {(1996,300) (1997,242) (1998,401) (1999,358) (2000,483) (2001,273) (2002,248) (2003,419) (2004,632) (2005,524) (2006,667) (2007,734) (2008,536) (2009,843) (2010,823) (2011,790) (2012,902) (2013,881) (2014,723) (2015,836) }
        node[left=.1cm,pos=0.7] {Articles};
        
    \end{axis}

    \begin{axis}[
        height=6cm,
        width=10cm,
        axis y line*=right,
        axis x line=none,
        ymin=0, ymax=100,
        xmin=1996, xmax=2015,
        ytick={0,20,40,60,80,100},
        ylabel={\textcolor{red}{Framing} and Public Approval},
        ylabel near ticks,
        yticklabel pos=right
    ]

    \addplot [very thick,red,mark=x]
        coordinates {(1996,40.3000000000000) (1997,40.5371900826446) (1998,42.0698254364090) (1999,44.8044692737430) (2000,43.1884057971015) (2001,43.0402930402930) (2002,41.4516129032258) (2003,43.3890214797136) (2004,42.7215189873418) (2005,27.8053435114504) (2006,29.6251874062969) (2007,27.9836512261580) (2008,25.9141791044776) (2009,22.5266903914591) (2010,23.3292831105711) (2011,26.7594936708861) (2012,28.5920177383592) (2013,26.6969353007946) (2014,26.1964038727524) (2015,28.3133971291866) }
        node[below right=.05cm,pos=0.65] {Negative Framing};

     \addplot [very thick,black,mark=x]
        coordinates {(1996,27) (1997,29) (1998,33) (1999,35) (2000,36) (2001,38) (2002,38) (2003,41) (2004,42) (2005,37) (2006,42) (2007,46) (2008,40) (2009,40) (2010,53) (2011,53) (2012,50) (2013,54) (2014,55) (2015,60) }
        node[below right=.05cm,pos=0.25] {Public Approval (Gallup)};

    \end{axis}
\end{tikzpicture}%
}

   
    \caption{News framing and public approval.}
    \label{fig-lgbt}
\end{figure}

The LGBT rights domain, depicted in Fig.~\ref{fig-lgbt} visually demonstrates the causal influence of framing on public opinion. In order to measure polarity of framing, we removed nouns from the news corpus, and used the average polarity \cite{sentiwordnet} of adjectives and adverbs within an annual corpus. Note that the negativity of framing drops in the 2004--2005 period, which coincides with a measured framing change from a manual survey\cite{lgbt}. The figure demonstrates an inverse relationship between framing negativity and public approval, as framing changed from one emphasizing morality to the current-day focus on equal rights.

Finally, we contribute the notion of framing concentration, measured by entropic news keywords. We use entropy between temporally disparate news corpora to rank individual n-grams for their utility in distinguishing the later corpus from the earlier one. Entropic keywords therefore represent the ``state" or ``concentration" of news at a given time. We use the number of keywords required to attain 97\% of dataset entropy (corresponding to three standard deviations of a normal distribution) to represent framing concentration. Fig.~\ref{fig-surveillance} depicts framing concentration versus time for the domain Surveillance, around the period of the Snowden revelations. Note that in 2013, a single n-gram (Snowden) suffices to represent 97\% of dataset entropy. We found causation between framing concentration, public attention, and legislation. 

\ks{make point about predictive utility}


\begin{figure}
    \centering
     \begin{tikzpicture}[%
        obj/.style={fill=gray!20,text height=0.3cm, minimum height=1.5cm,text centered,text width=2.4cm},
        node distance=1.0cm and 1.0cm, auto]
    
        \draw node [obj] (F) {News Framing};
        \draw node [below=of F] (dummy) {}; 
        \draw node [obj, below=of dummy] (N) {News Facts};
        \draw node [obj, right=of dummy] (O) {Public Opinion};
        \draw node [obj, right=of O] (L) {Legislation};
        
        \draw [->,ultra thick] (F.south) |- (O.west);
        \draw [->,ultra thick] (F.east) -| (L.north);
        \draw [->,ultra thick] (N.north) |- (O.west);
        \draw [->,ultra thick] (O)--(L);
    \end{tikzpicture}
    \caption{The causal path from news to legislation}
    \label{fig:influence}
\end{figure}
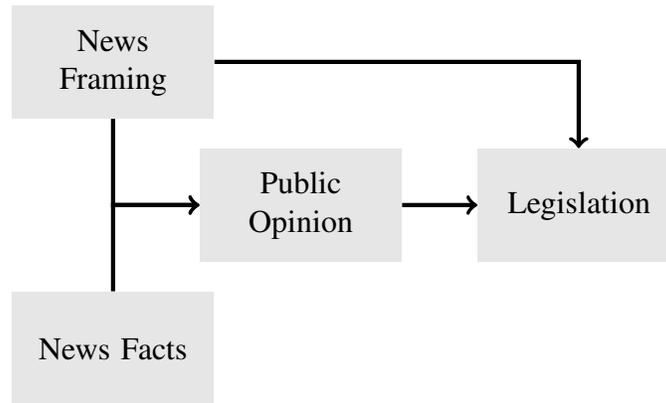

\begin{figure}[!htb]
    \centering
    \resizebox{0.99\columnwidth}{!}{%
\begin{tikzpicture}
    \begin{axis}[
        title={},
        height=6cm,
        width=10cm,
        xlabel={Year},
        ylabel={\textcolor{blue}{Article Count}},
        xmin=2004, xmax=2017,
        ymin=0, ymax=100,
        xtick={2004,2008,2012,2016},
        ytick={0,10,20,30,40,50,60,70,80,90,100},
        legend pos=north west,
        legend style={font=\tiny},
        mark size=2.0pt,
        ymajorgrids=true,
        grid style=dashed,
        xticklabel style={/pgf/number format/.cd, set thousands separator={}},
    ]

    \addplot [very thick,blue,mark=o]
        coordinates {(2004,1) (2005,0) (2006,1) (2007,0) (2008,2) (2009,1) (2010,2) (2011,6) (2012,17) (2013,59) (2014,59) (2015,74) (2016,78) (2017,69)}
        node[left=.1cm,pos=0.7] {Articles};

    \addplot [very thick,black,mark=square]
        coordinates {(2004,3.51851851851852) (2005,3.8889) (2006,3.88888888888889) (2007,4.4444) (2008,5.1852) (2009,8.5185) (2010,10.1852) (2011,12.9630) (2012,23.7037) (2013,52.2222) (2014,63.7037) (2015,99.6296) (2016,100) (2017,80)}
        node[left=.0cm,pos=0.7] {Google Trends};

    \end{axis}

    \begin{axis}[
        height=6cm,
        width=10cm,
        axis y line*=right,
        axis x line=none,
        ymin=0, ymax=1,
        xmin=2004, xmax=2017,
        ytick={0,0.20,0.40,0.60,0.80,1.00},
        ylabel={\textcolor{red}{Similarity} and GT index},
        ylabel near ticks,
        yticklabel pos=right
    ]

    \addplot [very thick,red,mark=x]
        coordinates {(2004,0) (2005,0) (2006,0.0322830578512397) (2007,0) (2008,0.0242122933884298) (2009,3.22830578512397e-04) (2010,0.00343677405445009) (2011,0.00942874162706814) (2012,0.230384688573092) (2013,0.231828700724021) (2014,0.263146275804372) (2015,0.292283420413794) (2016,0.372415638904476) (2017,0.312283420413063)}
        node[below right=.05cm,pos=0.65] {Similarity};
	
    \end{axis}
\end{tikzpicture}%
}
    \caption{News volume and correlation as predictors of legislation in Drones.}
    \label{fig-drones}
\end{figure}
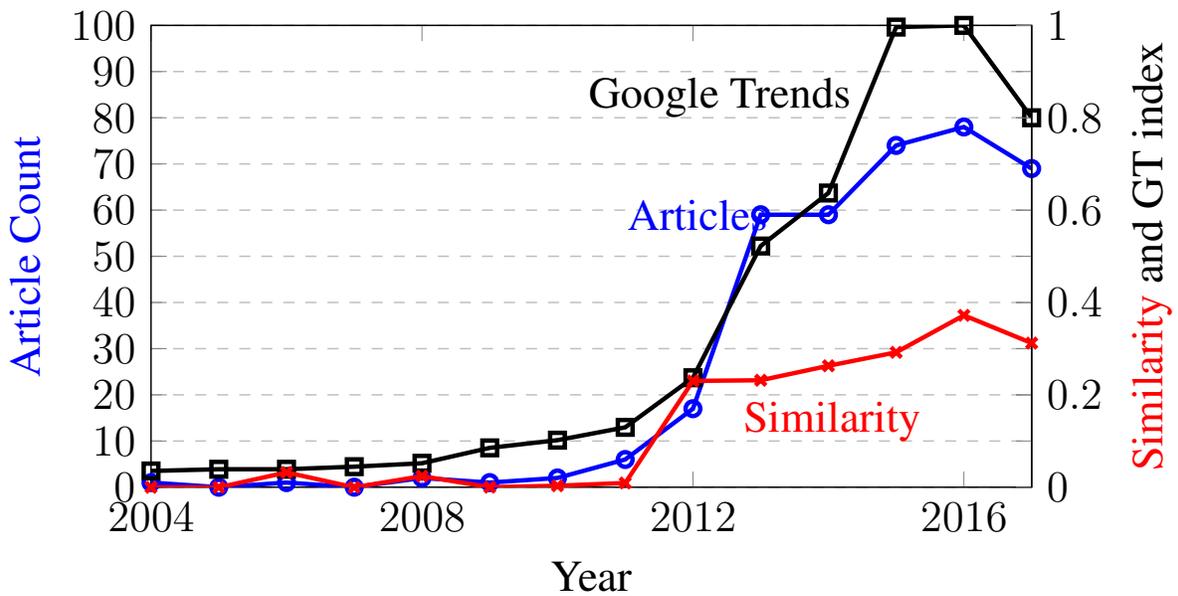

\begin{thebibliography}{10}

\bibitem{gunther1998persuasive}
A.~Gunther, {\it Communication Research\/} {\bf 25}, 486 (1998).

\bibitem{mutz1997reading}
D.~Mutz, J.~Soss, {\it Public Opinion Quarterly\/} pp. 431--451 (1997).

\bibitem{hoadley2010privacy}
C.~Hoadley, H.~Xu, J.~Lee, M.~B. Rosson, {\it Electronic Commerce Research and
  Applications\/}  (2010).

\bibitem{quora}
C.~Taylor, After privacy uproar, {Quora} feeds will no longer show data on what
  other users have viewed (2016). \url{https://goo.gl/9wG65R}.

\bibitem{agenda}
S.~Iyengar, D.~Kinder, {\it News that Matters: Television and American
  Opinion\/} (University of Chicago Press, 2010).

\bibitem{king}
G.~King, B.~Schneer, A.~White, {\it Science\/} {\bf 358}, 776 (2017).

\bibitem{granger}
C.~Granger, {\it Econometrica: Journal of the Econometric Society\/} pp.
  424--438 (1969).

\bibitem{entman}
R.~Entman, {\it Journal of Communication\/} {\bf 43}, 51 (1993).

\bibitem{nytapi}
NYT, {Developer API}s (2016). \url{http://developer.nytimes.com/}.

\bibitem{guardianapi}
{The Guardian}, {Guardian Open Platform},
  \url{http://open-platform.theguardian.com/} (2016). Accessed: 2016-3-3.

\bibitem{vg05}
A.~Viera, J.~M. Garrett, {\it Family Medicine 37, 5 (2005)\/} pp. 360--363
  (2005).

\bibitem{cd}
P.-N. Tan, {\it et~al.\/}, {\it Introduction to Data Mining\/} (Pearson
  Education India, 2006).

\bibitem{doc}
Q.~Le, T.~Mikolov, {\it Proceedings of the 31st International Conference on
  Machine Learning\/} (2014), pp. 1188--1196.

\bibitem{coppa}
FTC, {Children's Online Privacy Protection Rule} (1998).
  \url{https://www.ftc.gov/enforcement/rules/rulemaking-regulatory-reform-proceedings/childrens-online-privacy-protection-rule}.

\bibitem{ferpa}
{US Department of Education}, {Family Educational Rights and Privacy Act},
  \url{https://tinyurl.com/ybohwmfm} (1974).

\bibitem{gtapi}
P.~Trasborg, {The Google Trends API} (2018).
  \url{https://www.npmjs.com/package/google-trends-api}.

\bibitem{sentiwordnet}
S.~Baccianella, A.~Esuli, F.~Sebastiani, {\it Proceedings of the Seventh {ELRA}
  International Conference on Language Resources and Evaluation\/} (European
  Language Resources Association, Valletta, Malta, 2010).

\bibitem{lgbt}
S.~Engel, {\it Law and Social Inquiry\/} {\bf 38}, 403 (2013).
  \url{http://dx.doi.org/10.1111/j.1747-4469.2012.01319.x}.

\end{thebibliography}

\bibliographystyle{Science}

\end{document}